# Photothermal effects during nanodiamond synthesis from a carbon aerogel in a laser-heated diamond anvil cell

Matthew J. Crane[†], Bennett E. Smith[‡], Peter B. Meisenheimer[§], Xuezhe Zhou[§], Rhonda M. Stroud[^], E. James Davis[†], and Peter J. Pauzauskie[*, §, +]

† - Department of Chemical Engineering, University of Washington, Seattle WA 98195-1750

‡ - Department of Chemistry, University of Washington, Seattle, WA 98195-1700

§ - Department of Materials Science & Engineering, University of Washington, Seattle, WA 98195-2120

^ - Materials Science and Technology Division, Naval Research Laboratory, Washington, DC 20375

+ - Physical and Computational Sciences Directorate, Pacific Northwest National Laboratory

Richland, WA 99352

KEYWORDS: Diamond anvil cell, diamond point defects, photothermal heating, Mie theory, nitrogen vacancy

Corresponding Author. Tel: 206 543-2303. peterpz@uw.edu

**Abstract**

Nanodiamonds have emerged as promising materials for quantum computing, biolabeling, and sensing due to their ability to host color centers with remarkable photostability and long spin-coherence times at room temperature. Recently, a bottom-up, high-pressure, high-temperature (HPHT) approach was demonstrated for growing nanodiamonds with color centers from amorphous carbon precursors in a laser-heated diamond anvil cell (LH-DAC) that was supported by a near-hydrostatic noble gas pressure medium. However, a detailed understanding of the photothermal heating and its effect on diamond growth, including the phase conversion conditions and the temperature-dependence of color center formation, has not been reported. In this work, we measure blackbody radiation during LH-DAC synthesis of nanodiamond from carbon aerogel to examine these temperature-dependent effects. Blackbody temperature measurements suggest that nanodiamond growth can occur at 16.3 GPa and 1800 K. We use Mie theory and analytical heat transport to develop a predictive photothermal heating model. This model demonstrates that melting the noble gas pressure medium during laser heating decreases the local thermal conductivity to drive a high spatial resolution of phase conversion to diamond. Finally, we observe a temperature-dependent formation of nitrogen vacancy centers and interpret this phenomenon in the context of HPHT carbon vacancy diffusion using CBΩ theory.

1. **Introduction**

Nanodiamond materials have become the subject of interest for a wide range of optical and electronic applications due to their ability to contain a variety of multifunctional color centers. A fundamental understanding of point-defects within diamond has led to numerous intriguing applications including optically-initialized quantum bits for quantum computing via spin polarization, high fidelity sensing of the local molecular environment, and bright fluorescence for sub-diffraction-limit imaging [1–3]. Because nanodiamond is biocompatible, these can all be accomplished in *in vitro* or *in vivo* with nanometer-scale resolution. Additional applications include catalysis, composites, and drug delivery [4–6]. However, the ability to deterministically introduce heteroatomic defects (e.g. the NE8 center) remains limited, and the pursuit of heteroatomic defects typically relies on ion implantation, diffusion, and confocal imaging to search for the desired color centers [7–9].

The most common synthetic strategy to produce nanodiamonds employs dynamic (< 10 ns) high-pressure, high-temperature (HPHT) conditions to nucleate nanodiamond grains. For example, detonation nanodiamonds form within a shockwave produced by the confined detonation of a high explosive [10]. As the shockwave propagates, the supersaturated carbon vapor rapidly condenses into nanoscale liquid droplets that homogeneously nucleate nanocrystalline diamonds. Ultrasound cavitation, laser ablation, and shock waves all produce similar conditions and lead to nanodiamond formation [11–13]. The doping of nanodiamonds synthesized with these dynamic processes occurs by varying the chemical composition and bonding of the carbon precursor, which provides a potential route to influence the formation of optically-active, heteroatomic point-defects [14,15]. However, the brief duration of the high-temperatures generated in these methods produce a wide range of lattice defects and unintentional incorporation of heteroatoms from the synthesis chamber.

This lack of synthetic control ultimately limits the optoelectronic application of detonation nanodiamond materials [16].

Bulk diamond films grown via chemical vapor deposition (CVD) exhibit significantly greater chemical purity and electronic quality in comparison with detonation nanodiamond. In addition, ion-irradiation of these films enables the incorporation of heteroatoms. Ion sources can achieve a high degree of spatial dopant control through the use of masks and SRIM calculations [17]. Subsequent thermal annealing and acid-washing can reduce lattice damage, remove $sp^2$ carbon, and drive carbon-vacancy ($V_c$) diffusion to generate optically-active defects [8,17]. In addition, recent advances with pulsed lasers have enabled a high precision of $V_c$ incorporation to potentially replace this annealing step [18]. Ball milling or etching of CVD-synthesized diamond films can produce nanodiamonds via a top-down approach with similarly high quality as the original films, as evidenced by the long $T_2^*$ lifetimes in the resulting nanodiamonds [19]. However, it has remained a challenge to use CVD methods to create complex, polyatomic defect centers in diamond, due to the decomposition of heteroatomic precursors in the extreme non-equilibrium plasma environment.

The static, HPHT synthesis of diamond also can yield bulk crystals with electronic properties similar to ion-implanted CVD diamond. In fact, during the first successful HPHT experiments in the 1960's, nanodiamonds were an unexpected byproduct during the first successful production of bulk diamond at an industrial scale. The direct HPHT synthesis of nanodiamonds was not pursued again until recently [20–24]. Research on bulk HPHT diamond during this 50-year hiatus realized a range of important discoveries including the use of catalysts to decrease the pressures and temperatures required for diamond production, the observation that different carbon precursors can form diamond at less extreme conditions, and the ability to incorporate dopants into diamond by simply mixing them into the carbon precursor [20,22,25–28].

Recently, nanodiamonds were synthesized at HPHT in a LH-DAC from a carbon aerogel precursor, a nanostructured sol-gel of amorphous carbon aerogel that is produced with solution-phase chemistry [20]. In this approach, the absorption of a near-infrared laser increased the temperature of the amorphous carbon aerogel target at high (>20 GPa) pressures to drive a phase transition from amorphous carbon to diamond nanocrystals. The authors hypothesized that the use of a carbon aerogel precursor confined diamond nucleation and growth to yield a nanocrystalline product. Intriguingly, sol-gel chemistry provides many degrees of freedom for precise heteroatomic doping beyond what is possible in ion implantation of CVD diamond. However, a number of challenges remain in understanding the laser-driven HPHT formation of diamond including the kinetics of photothermal heating, the interplay between temperature and atomic diffusion during growth, and also the potential for Ostwald ripening of nanocrystalline grains during extreme photothermal heating [24,25,29] to temperatures approaching 2,000K.

In this paper, we present the first predictive photothermal model of HPHT nanodiamond synthesis from amorphous carbon aerogel in a laser-heated DAC. Using Mie theory, we develop an analytical model of heat transport to quantify the maximum temperatures during nanodiamond formation. By correlating temperatures during phase conversion with Raman and photoluminescence, we identify that the nanodiamond-carbon aerogel phase boundary sits between 12 and 16 GPa and 1600 to 1800 K. This is significantly below the graphite-diamond boundary [30]. Analytical modeling and irradiance-dependent blackbody (Planck) emission show that the solid argon pressure medium melts to form a thermally-insulating, supercritical fluid during the synthesis. In addition, the photoluminescence of nanodiamond synthesized at the lowest laser irradiance does not exhibit NV⁻ centers, while PL at the highest temperatures contains NV⁻ defects. A temperature- and pressure-dependent CBΩ model for vacancy diffusion suggests that NV formation depends on $V_c$ diffusion to substitutional nitrogen atoms, which is modified at HPHT conditions.

## 2. Experimental

### 2.1 Aerogel synthesis

Carbon aerogels were synthesized by dissolving resorcinol (Sigma-Aldrich) in acetonitrile (EDM Millipore) and adding formaldehyde (37 wt. % in $H_2O$ with methanol stabilizer, Sigma-Aldrich), followed immediately by hydrochloric acid catalyst (37 wt. % Macron Fine Chemicals) [20]. The final molar ratios of acetonitrile, formaldehyde, and HCl relative to resorcinol were 100:1, 2:1, and 1:10, respectively [31,32]. This solution was quickly placed into a Branson 1510R-DTH ultrasonic cleaner for 30 minutes, and then removed and allowed to gel for about 24 hours, during which the gel turned light pink. The acetonitrile solvent was exchanged by washing with ethanol 4 times over 5 days. Aerogels were then dried in an autoclave filled (E3100, Quorum Technologies) with supercritical $CO_2$ to prevent collapse of the pore structure. Finally, the gels were pyrolyzed under $N_2$ at 900°C for 4 hours to remove oxygen species. The gel surface area was measured with a NOVA 2200e Surface Area and Pore Size Analyzer.

### 2.2 Diamond anvil cell loading, heating, and spectroscopy

A Bohler-Almax plate DAC with 0.300 mm diameter culets was prepared by dimpling a rhenium gasket to a thickness of 30 – 40 μm and drilling a ~120um hole. Thickness was measured using either Fabry-Pérot interference fringes or a calibrated micrometer. The gasket, which acts as the walls of the pressure chamber, was then carefully aligned into the DAC, which acts as the roof and floor of the pressure chamber. Gels were lightly ground in aluminum foil and transferred into the gasket with a Mäerzhäuser Wetzlar nanomanipulator. Gaseous argon was condensed with liquid $N_2$ and added to the DAC as a near-hydrostatic pressure medium. The liquid argon infiltrates the aerogel's pores during this process, allowing the gel to maintain its morphology throughout the phase transformation to diamond [20]. The DAC pressure was incrementally increased to a final pressure of 16.3 GPa as shown in Fig. S5. Additional experiments were performed at 21 GPa. Pressure was measured *in situ* by exciting a ruby crystal placed in the cell and measuring the $^2T_1$ to $^4A_2$ $Cr^{3+}$ emission (Fig. S5), which has a well-documented pressure dependence [33]. During compression to 16.3 GPa, the argon pressure medium undergoes a phase transition to solid, face-centered cubic argon [34]. In addition, higher laser irradiances were chosen to ensure similar synthesis temperatures.

Laser heating in the DAC was accomplished with a solid-state Nd:YAG (BL-106C, Spectra Physics) focused via a Mitutoyo 50x objective (0.55 NA) using a polarized beam splitter cube. Raman and PL were collected with a home-built setup comprised of a Coherent compass 532 nm laser, the aforementioned Mitutoyo objective, and focused onto an Acton SpectraPro 500i spectrometer with a Princeton Instruments liquid $N_2$-cooled CCD detector. Thermal Planck emission was collected with an Ocean Optics USB2000 spectrometer. These spectra were fit to Wien's approximation, and the result of this linear fit was used as the starting parameter for a nonlinear fit to Planck's law. All spectra were collected and intensity-corrected with an Ocean Optics HL-2000 intensity calibration lamp and wavelength-corrected with an argon lamp. The analysis of electromagnetic heating and its implementation with Python are provided in the Supplementary Information.

Transmission electron microscope (TEM) bright field images of the aerogel precursor and HPHT products were taken with an FEI Tecnai G2 F20 at an accelerating voltage of 200 kV. Selected area electron diffraction (SAED) patterns of the products were obtained with a JEOL 2200FS at the Naval Research Lab (NRL), and analyzed with calibration constants derived from aluminum diffraction standards. Aberration-corrected scanning transmission electron microscopy (STEM) studies were performed at 60 kV with a Nion UltraSTEM 200-X at NRL. The NRL UltraSTEM is equipped with a Gatan Enfinium ER electron energy loss spectrometer and Bruker energy dispersive X-ray spectrometer for electron energy loss spectroscopy (EELS) and energy dispersive X-ray spectroscopy (EDS) measurements, respectively.

3. **Results and discussion**

*3.1 Nanodiamond synthesis and patterning*

To study photothermal effects during nanodiamond growth, we synthesized an amorphous carbon aerogel, which was subsequently pressurized and laser heated in a DAC [35]. The carbon aerogel consisted of an interconnected, three-dimensional network of amorphous carbon spheres (12.9 ± 3.4 nm diameter, Fig. 4a) with a moderate surface area (311.6 $m^2$/g from nitrogen sorption), low thermal conductivity (approximately $10^{-2}$ W/m-K [36]), and broadband spectral absorption [37]. To examine the effects of temperature on the nanodiamond product, we focused an infrared, continuous-wave 1064 nm laser onto the high-pressure carbon aerogel and collected Planck emission, moving to new areas before each measurement. While increasing the heating irradiance to 17 MW/$cm^2$, the average carbon aerogel temperature increased monotonically up to 1340 K (Fig. 2), below the melting point of argon (1383 K at 16.3 GPa [34]). At these irradiances, the temperature remained relatively constant throughout heating (within one standard deviation of the average temperature), and no nanodiamond formation was observed. This was consistent with previous reports of HPHT synthesis, which did not report diamond formation below 1600 K at 21 GPa or 2300 K at 15 GPa (Fig. 1a) [22,25,27,30]. Above 17 MW/$cm^2$, the temperature rapidly increased above 1800 K, greater than the melting point of argon, for 1-5 seconds before decreasing to 1400 K for the remainder of the laser-heating, as shown in Fig. 2a, and nanodiamond formation was observed. In addition, the maximum synthesis temperature remained constant with increasing laser irradiance. These data suggest that argon's high thermal conductivity limits nanodiamond formation and that the laser-heated HPHT nanodiamond synthesis only occurs after the pressure medium melts, decreasing its thermal conductivity. In addition, the constant maximum synthesis temperature implies that the phase conversion of carbon aerogel to nanodiamond self-limits further heating.

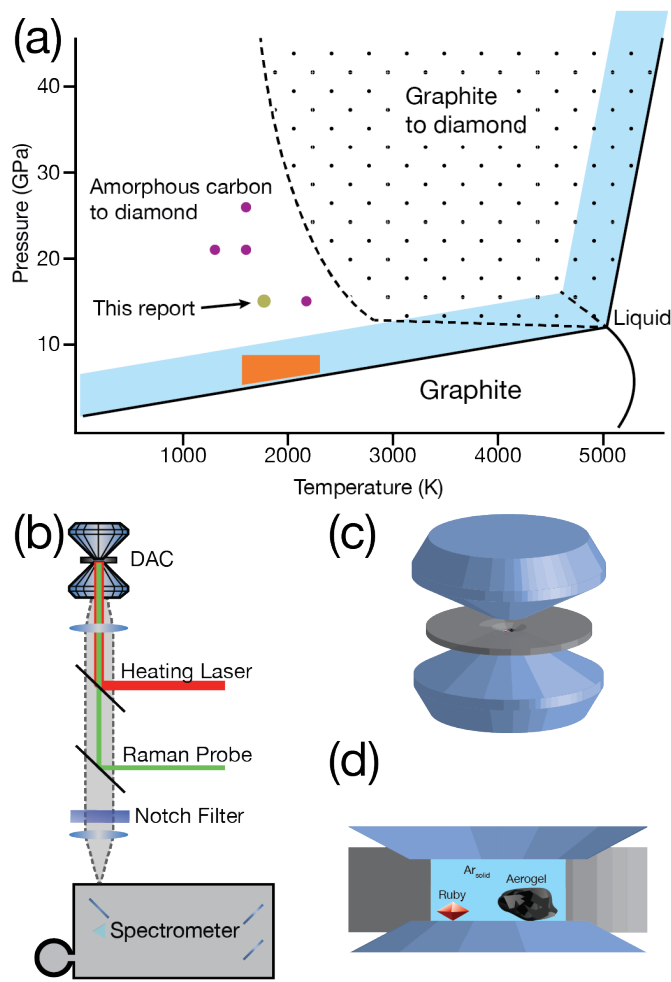

**Figure 1** Diamond-graphite phase diagram with kinetically-accessible regions (a). The dotted region labeled shows the direct graphite-to-diamond phase transition. The light cyan region represents the phase diagram for a 1.2 nm diameter diamond nanocrystal, and the orange area represents the experimental region for catalyzed graphite-to-diamond phase conversion. The dark purple dots represent the successful synthesis of diamond from amorphous, non-graphite carbon starting materials. A schematic of spectroscopy system for laser-heating and Raman experiments (b) and the laser-heated DAC (c), with an expanded view of ruby and aerogel under hydrostatic compression by solid argon (d).

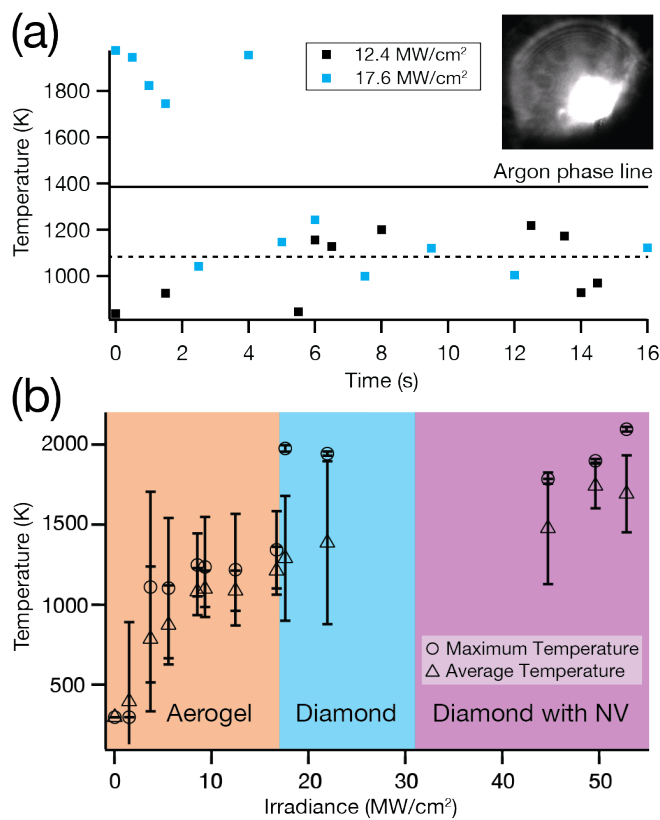

**Figure 2** Time-dependent, Planck-fit temperatures during laser irradiation of carbon aerogel at a pressure of 16.3 GPa at 17.6 MW/cm$^2$, which melts argon, and 12.4 MW/cm$^2$, which does not melt argon (a). The dotted line represents the average temperature at 12.4 MW/cm$^2$ and the solid line

represents the argon solid-liquid phase line. The inset shows an image captured during laser heating. The time-averaged and maximum temperatures measured during the laser heating of carbon aerogel inside the DAC at different irradiances (b). The orange, blue, and purple panels represent the phase space where the aerogel did not undergo a phase change, where cubic diamond formed without NV, and where cubic diamond formed with NV defects, respectively. The average temperature error bars represent the standard deviation from at least 60 spectra collected over a 30 s period; the maximum temperature error bars represent 95% confidence interval fits to Planck's Law.

Upon heating above 1800 K, transparent regions appeared in the carbon aerogel with new Raman scattering modes—480, 1130, 1370, 1410 and 1550 $cm^{-1}$—corresponding to the formation of diamond coated with amorphous surface material, graphite, and transpolyacetylene, which are common in HPHT diamond synthesis (Fig. 3c, Fig. 3d, and Fig. S1) [20,22,25,38,39]. We distinguished potential photoluminescence (PL) from Raman scattering by observing peak shifts for Raman modes with varying excitation laser wavelength (Fig. S1). Selected-area electron diffraction (SAED), TEM images, and EELS spectra of the carbon aerogel after HPHT processing showed crystalline structures composed of $sp^3$ carbon with a small amount of $sp^2$ carbon and a 2.06 Å lattice spacing, corresponding to nanodiamond with $sp^2$ surface reconstruction (Fig. 4). The EELS spectra also contain a dip at 302.5 eV, characteristic of cubic diamond's second gap. The minimum conditions (16.3 GPa and ~1800 K) to drive the phase change of carbon aerogel to nanodiamond are among the lowest reported, as illustrated in Figure 1a [22,25,27,40]. In a separate experiment at 12.0 GPa and identical irradiances, there was no evidence of phase conversion, suggesting that the carbon aerogel–nanodiamond phase line sits between 12.0 and 16.3 GPa and 1340 and 1800 K. These relatively mild conditions are likely due to the amorphous carbon starting material and the rapid heating conditions. Amorphous carbon lacks long-range crystalline structure, which lowers the kinetic barrier present between other carbon allotropes like graphite (~0.4 eV) and diamond, even at high pressure [41,42]. The rapid heating of the carbon aerogel upon melting of the argon pressure medium may prevent graphitization that occurs in resistively-heated HPHT experiments prior to the formation of cubic diamond, thereby lowering the pressures and temperatures required for phase conversion [25,40].

This rapid heating may also enable the observation of new allotropes of carbon and will be the focus of future studies. However, this also suggests that the thermal gradients during photothermal heating due to the variation of the irradiance within the Gaussian laser (or the diffusion of heat away from the beam spot) may influence phase conversion to diamond. Indeed, while most recovered material predominantly contained nanodiamond, some grains exhibited greater concentrations of graphite accompanied by a small amount of carbon onions. Carbon onions were not observed in recovered material with large concentrations of nanodiamond (Fig. 4b). These data suggest that thermal gradients exist during the synthesis and that $sp^2$ domains likely form in lower temperature regions. Because the synthesis conditions are insufficient to drive phase conversion from graphite to nanodiamond, graphite represents a kinetic dead end and cannot produce diamond during subsequent direct photothermal heating.

Interestingly, the large thermal gradients due to the high thermal conductivity of solid argon and low thermal conductivity of carbon aerogel resulted in localized heating that limited phase conversion to a small area (Fig. 3 and Fig. S7). By varying the irradiance, the carbon-aerogel could be heated to a temperature greater than 1800 K. Heating at 47 MW/cm$^2$ and rastering over the aerogel produced a diamond line-width of 8.5 ± 0.3 μm. Heating and rastering at 22 MW/cm$^2$ produced a diamond line width of 4.0 ± 0.6 μm. To the best of the author's knowledge, this is the highest resolution of diamond film patterning to date [43].

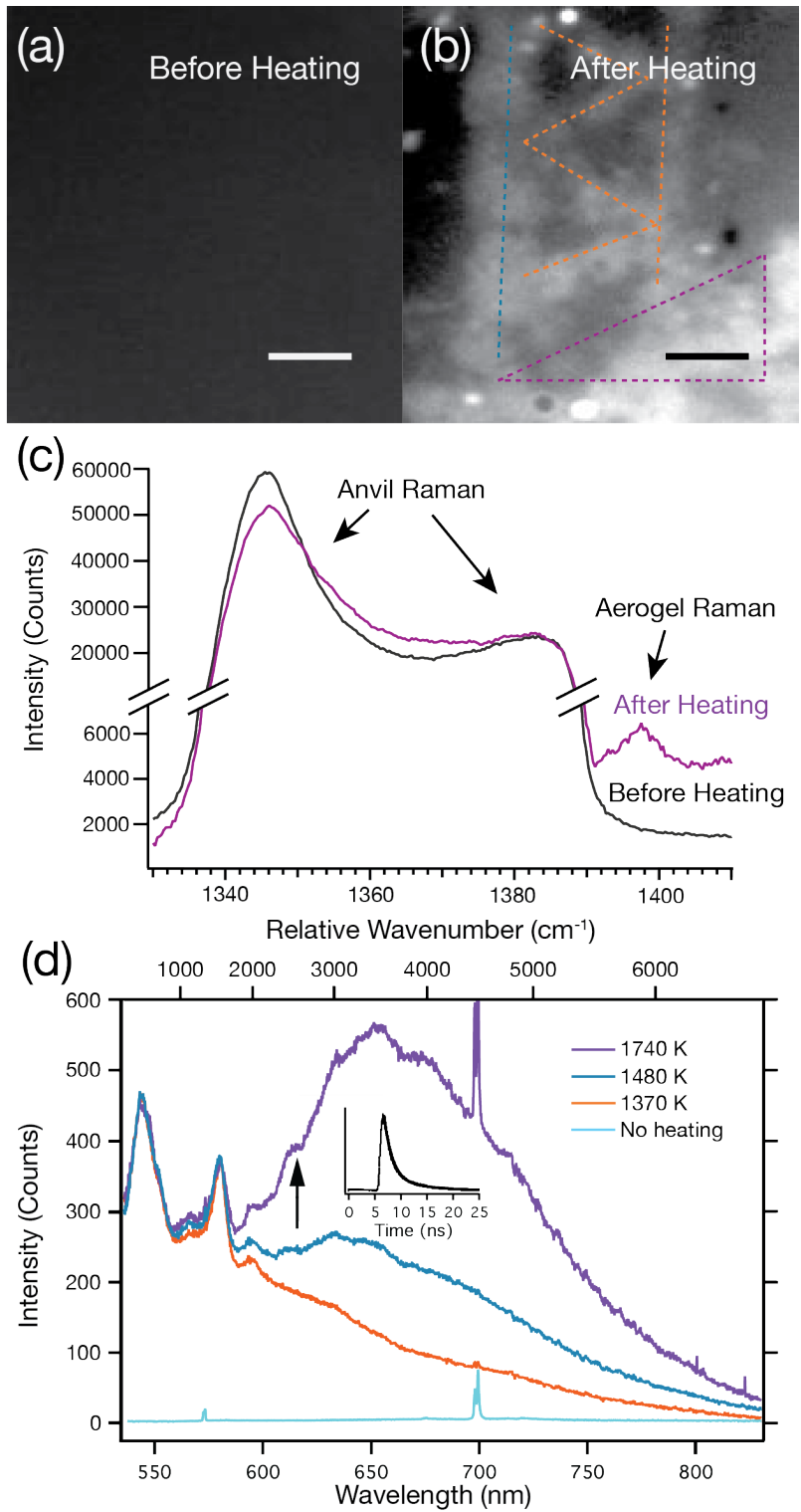

**Figure 3** Dark-field optical micrographs before (a) and after (b) aerogel heating in the DAC at 16.3 GPa, respectively. Heating on the left side of panel b at 47 MW/cm$^2$ yields a diamond line-width of 8.5 ± 0.3 μm (orange dotted line); heating on the right side of panel b at 22 MW/cm$^2$ (blue dotted line). The bottom right corner shows a large area of heating at 53 MW/cm$^2$ (purple dotted line). Scale bars in both images are 10 μm. Raman from the heated areas (c) and characteristic emission (d) from diamond after subsequent heating and excitation at 1 mW and 532 nm normalized to the graphite g-band. The black arrow shows the NV$^-$ ZPL at 16.3 GPa and the decay inset shows time-correlated single photon counting at this wavelength. These data are fit with two time constants at 1.2 and 7.0 ns. The sharp peaks centered at 700 nm is due to the $^2T_1$ to $^4A_2$ crystal-field transition of $Cr^{3+}$ in the ruby pressure sensors.

*3.2 High pressure defect formation*

In previous reports of HPHT nanodiamond from a carbon aerogel, SiV$^-$ defects with a ZPL photoluminescence peak at 738 nm were observed [20]; however, in these experiments, SiV$^-$ emission was not observed at any conditions. These previous HPHT diamond aerogel experiments used a carbon aerogel precursor that was synthesized under mildly *basic*, aqueous conditions (pH~8.4) using silica glassware; whereas, in the experiments reported here, we carefully avoided silica glassware and employed an alternative *acidic* catalyst in a non-aqueous organic solvent (acetonitrile). These observations suggest that silicon incorporates into the carbon grains as silicic acid during polymerization by basic etching of the silica glassware and will be the focus of future investigations [44].

The photoluminescence (PL) spectra of nanodiamonds synthesized at average temperatures above 1480 K showed a zero phonon line (ZPL) at 610, phonon sidebands centered at 650 nm, and a lifetime of 8.2 ns, characteristic of an NV$^-$ center under hydrostatic pressure at 16.3 GPa (Fig.

3). This is consistent with prior reports at pressure [45,46]. Surprisingly, NV⁻ production was not observed in nanodiamond formed with an average temperature up to 1340 K. Because the diamond formation temperatures are self-limiting and constant (1950 K average) irrespective of irradiance, the temperature-dependence of NV⁻ formation is likely not due to size or surface effects. Electron paramagnetic resonance, optical absorption, and PL spectroscopy studies have shown that nitrogen initially incorporates in diamond as a substitutional ($N_s$) defect, which is immobile at pressures above 6 GPa for temperatures <2000 K [47]. Comparably, neutral carbon vacancy defects in diamond are highly mobile, and they diffuse upon annealing until they meet $N_s$, where they ultimately form low-energy NV⁻ defects. Annealing-induced vacancy-diffusion is commonly used to generate NV⁻ defects, and the temperature dependence of this process at atmospheric pressure is well documented [48,49]. However, the neutral vacancy diffusion coefficient decreases significantly with increasing pressure due to compression of the diamond lattice. In addition, while the carbon aerogel in this work was laser-heated in 30 second intervals, diamond films are typically annealed for ~2 hours to produce NV⁻ centers. Using CBΩ theory to model vacancy diffusion at 16.3 GPa and a probability distribution function solution for 3D diffusion, we estimate that vacancies will consistently activate NV⁻ defects at ~1350 K (Table 1, Fig. S3) [47,50,51]. This boundary is consistent with the observed temperature range for NV⁻ formation, considering potential variations in $N_s$ concentration. These observations suggest that the formation of NV⁻ centers in laser heated diamond occurs via the diffusion of carbon vacancies to nitrogen, which is initially incorporated as $N_s$ below 1 wt-%, and that we can accurately predict these conditions [52,53]. Interestingly, this also implies that varying the pressure could provide the ability to control defect formation. Finally, while there have been recent reports on strain-alignment of defects, we do not anticipate any preferential alignment due to the near-hydrostatic nature of the argon pressure medium [54].

**Table 1.** Observed and calculated regions of NV formation from cBΩ theory

|  | At 1 atm[*] (K) | Observed (K) | At 16.3 GPa (K) |
| --- | --- | --- | --- |
| Vacancies become mobile | 870 | - | 1070 |
| NV Centers begin to form | 1070 | 1480 | 1350 |
| NV Centers with bright photoluminescence | 1470 | 1700 | 1950 |

[*]Ref. [48]

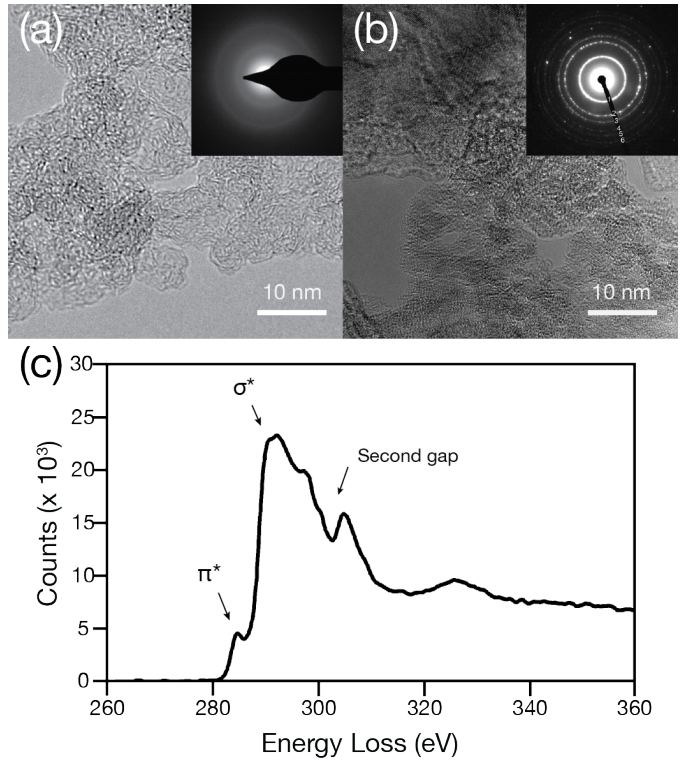

**Figure 4** Bright field TEM images of carbon aerogel before (a) and after (b) HPHT treatment and the electron energy loss spectrum (c) of the HPHT treated carbon aerogel. The selected area electron diffraction (inset) before HPHT treatment shows amorphous material, while the pattern after HPHT treatment indexes to cubic diamond. The labels 1-6 correspond to the measured values and (hkl) indices of (1) 2.06 Å (111); (2) 1.29 Å (220); (3) 1.09 Å (311); (4) 0.92 Å; (5) 0.83 Å (331); and (6) 0.75 Å (422).

*3.3 Photothermal heating model*

To better understand the nanodiamond formation within a LH-DAC and quantify the ultimate resolution of phase conversion, we developed a predictive analytical heat transport model for single amorphous-carbon spheres under irradiation from a continuous-wave laser source. Based on the analysis of TEM images (Fig. 4a), we modeled the carbon aerogel as spheres with 12.9 nm diameters. It is important to note that the resolution of the phase transformation depends not on the spot size of the laser, but rather on the ability of the amorphous carbon precursor to support large internal temperature gradients [55]. Thus, the temperature distribution, which is a function of the pressure medium, carbon sphere, and heating laser dictate the ultimate resolution of diamond synthesis and dopant formation. To solve for the temperature distribution, we apply Mie theory to obtain the internal electric field [56] and use this solution as the source term in the energy balance, given by

$$\rho C_p \frac{\partial T}{\partial t} = \frac{1}{r^2}\frac{\partial}{\partial r}\left(kr^2 \frac{\partial T}{\partial r}\right) + \frac{1}{r^2 \sin\theta}\frac{\partial}{\partial \theta}\left(k \sin\theta \frac{\partial T}{\partial \theta}\right) + \frac{1}{r^2 \sin^2\theta}\frac{\partial}{\partial \phi}\left(k \frac{\partial T}{\partial \phi}\right) \quad (1)$$
$$+ S(r,\theta,\phi),$$

where $T$ is the temperature, $r$, $\theta$, and $\phi$ are the radial coordinate, polar angle, and azimuthal angle defined in Fig. 5c, and $k$, $\rho$, and $C_p$ are the thermal conductivity, density, and heat capacity of the carbon sphere. For highly absorbing spheres in the Rayleigh limit, the source function, $S$, is only a function of $r$ [56]. Consequently, at steady state the internal temperature is only a function of $r$, and the one-dimensional energy equation reduces to (for constant thermal conductivity evaluated at the mean temperature of the sample)

$$\frac{d}{dr}\left(r^2 \frac{dT}{dr}\right) = -\frac{S}{k} r^2 \qquad (2)$$

The temperature is bounded at $r = 0$, and the boundary condition at the surface is a combination of conductive and radiative heat losses given by

$$-k \frac{dT}{dr}\bigg|_{r=R} = \frac{k_{Ar}}{R}(T_R - T_\infty) + \sigma(T_R^4 - T_\infty^4), \qquad (3)$$

where $T_\infty$ is the bulk mean temperature of the surroundings, $\sigma$ is the Stefan-Boltzmann constant, and $k_{Ar}$ is the argon thermal conductivity evaluated at the surface temperature, $T_R$. The radiation may be modified by writing

$$T_R^4 - T_\infty^4 = (T_R + T_\infty)(T_R^2 + T_\infty^2)(T_R - T_\infty). \qquad (4)$$

The solution of the energy equation becomes

$$T = T_\infty + \frac{SR^2}{3k}\left(\frac{1}{Bi} + \frac{1}{2}\left(1 - \left(\frac{r}{R}\right)^2\right)\right), \qquad (5)$$

where the Biot number is defined by

$$Bi = \frac{k_{Ar}}{k} + \frac{\sigma R}{k}(T_R + T_\infty)(T_R^2 + T_\infty^2). \qquad (6)$$

The surface temperature is given by

$$T_R = T_\infty + \frac{SR^2}{3kBi}, \qquad (7)$$

which is determined by solving equations (5) and (6) simultaneously. The results of this model provide several important insights. First, the temperatures achieved within carbon grains depend heavily on the thermal conductivity of the argon pressure medium, which melts to form supercritical argon during photothermal heating. The thermal conductivity of supercritical argon is challenging to measure at these conditions. We bound the thermal conductivities by using extrapolated, temperature-dependent gaseous (1 atm) and solid (16.3 GPa) argon properties to calculate the maximum and minimum temperatures, respectively. This analysis suggests the maximum internal temperature to be between 1750 and 2200 K during photothermal heating at 18 MW/cm$^2$ (Fig. 5c). However, based on the experimental conditions, we expect the supercritical argon to exhibit gaseous characteristics and anticipate that temperatures will be closer to 2200 K, which is consistent with the observed temperatures [57,58]. Using this assumption, the photothermal heating of isolated carbon spheres with diameters above 10 nm can drive a phase transition to diamond in supercritical argon (Table S1). Below 10 nm, both collective absorption and lateral heat insulation are required to produce diamond. Synthetic control over the monodispersity of carbon spheres above 10 nm would enable the most precise resolution of nanodiamond formation by selectively converting material at the center of the focused, laser beam [55]. Similarly, leveraging the thermal conductivity of the pressure medium through the selection of either different pressure media or different pressure ranges may enable tuning of this size range in future experiments.

Second, there exists a large temperature gradient throughout the individual carbon grains during photothermal heating, and the maximum occurs in the center of the carbon spheres (Fig. 5). At this pressure, diamond nucleates and grows, rather than forming via a diffusionless mechanism. We hypothesize that diamond nucleates at the sphere's center and growth continues outward, until the decrease in absorption due to phase conversion of amorphous carbon (n=1.5-1.0i) to diamond (n=2.4-0.0i) inhibits further conversion [41]. This process traps $N_s$ defects, which can later be converted to NV⁻, and coats the surface in the graphitic or amorphous carbon found in Raman scattering and EELS spectroscopy.

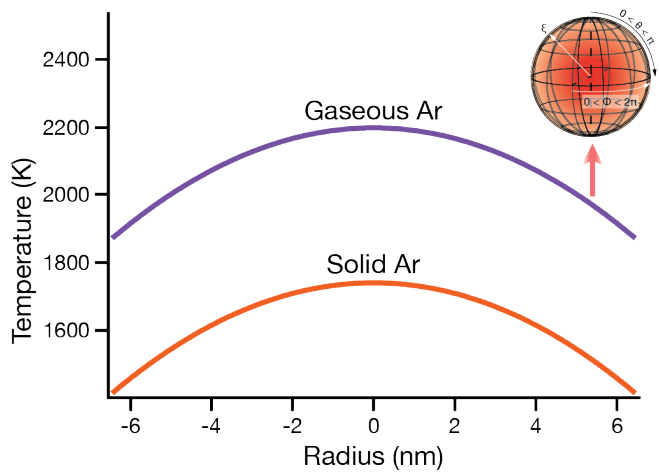

**Figure 5** Temperatures calculated within a single amorphous carbon sphere with a refractive index of 1.5 − 1.0i and thermal conductivity of 0.05 W/mK in supercritical argon (refractive index 1.6) at 18 MW/cm² for solid and gaseous argon (c). To account for sphere-sphere heating, we set $T_\infty$ to 1380 K, the melting point of argon at 16.3 GPa. Based on analysis of TEM images (Fig. 4a), we modeled the carbon aerogel as spheres with 12.9 nm diameters. The iterative solutions for gaseous and solid argon yielded thermal conductivities of 0.066 W/mK and 0.88 W/mK, respectively. The inset shows the coordinate system for temperature calculations within a single amorphous carbon sphere upon irradiation from the bottom of the sphere.

4. **Conclusions**

Progress in the development of nanodiamond materials for quantum computing and sensing applications increasingly relies on improving synthetic methods to enable precise control in the formation of point defects in the nanodiamond's crystal lattice. Experimental and theoretical results presented here show that nanodiamond formation from carbon aerogel can occur at 1800 K and 16.3 GPa. A combination of Mie theory and analytical heat transport predict that nanodiamond synthesis occurs after the argon pressure medium melts to form a supercritical fluid with low thermal conductivity. The thermal insulation from the supercritical argon causes rapid local heating of the carbon aerogel, likely leading to nucleation at the center of amorphous carbon grains, and subsequent radial growth until conversion to diamond limits further heating. Eliminating glassware during sol-gel synthesis suppressed the formation of SiV$^-$ centers. In addition, the temperature-dependent formation of NV$^-$ centers and CBΩ model suggest that nitrogen initial incorporates substitutionally and subsequent vacancy diffusion drives the formation of NV$^-$. The low thermal conductivity of carbon aerogel enables the highest reported resolution of phase conversion of amorphous carbon to diamond with a spatial resolution of 4 μm. The methodology presented here opens the door to diamond synthesis below the diffraction limit through high-resolution photothermal phase control [55].

*Electronic Supporting Information*

The following files are available:

Wavelength-dependent luminescence to separate photoluminescence from Raman scattering, additional spectra from NV- centers, a derivation of the vacancy diffusion model, a description of the Mie-Grüneisen-Debye model for the diamond equation of state, a comparison of the atmospheric pressure and high pressure vacancy diffusion coefficients in diamond, the full derivation of the sphere heating theory, all parameters used for modeling in the article, ruby pressure measurements, energy dispersive x-ray spectroscopy of the aerogel, and images of laser heating within the DAC.

*Author Contributions*

M.J.C. wrote the manuscript, conducted and designed the experiments, and implemented heat transport theory. B.E.S. assisted with DAC alignment and sample preparation. P.B.M. synthesized aerogels, assisted with DAC sample preparation, and implemented heat transport theory. X.Z. acquired TEM data. R.M.S. performed STEM-EELS and diffraction analysis. E.J.D. contributed analysis of heat transport and Mie theory. P.J.P. conceived of and directed the experiments and data analysis. All authors contributed to writing the manuscript.

*Funding Sources*

CAREER Award from the National Science Foundation (Award #1555007), startup funding from the University of Washington, a capital equipment donation from Lawrence Livermore National Lab, and Department of Defense National Science and Engineering Graduate (NDSEG) program.

*Acknowledgements*

This research was made possible by a CAREER Award from the National Science Foundation (Award #1555007), startup funding from the University of Washington, as well as a capital equipment donation from Lawrence Livermore National Lab. In addition, P.J.P. gratefully acknowledges support from both the US Department of Energy's Pacific Northwest National Laboratory (PNNL) and the Materials Synthesis and Simulation Across Scales (MS3) Initiative, a Laboratory Directed Research and Development (LDRD) program at the PNNL. The PNNL is operated by Battelle under Contract DE-AC05-76RL01830. M.J.C. gratefully acknowledges support from the Department of Defense through a National Defense Science and Engineering Graduate Fellowship (NDSEG) program and the Microanalysis Society through a Joseph Goldstein Scholar Award. The authors would like to thank Evan Abramson for his assistance loading and aligning the DAC and conversations regarding laser heating. The authors would also like to thank Xiaodong Xu and Kyle Seyler for their help with TCSPC and photoluminescence excitation spectroscopy experiments, and Chris Mundy for helpful discussions about molecular dynamic modeling of argon thermal conductivity.